\newcommand{\CLASSINPUTtoptextmargin}{1.4cm}
\newcommand{\CLASSINPUTbottomtextmargin}{1.4cm}
\newcommand{\CLASSINPUTinnersidemargin}{1.4cm}
\newcommand{\CLASSINPUToutersidemargin}{1.4cm}
\documentclass[conference,a4paper]{IEEEtran}
\IEEEoverridecommandlockouts
\usepackage{url}
\usepackage{subcaption}
\usepackage{cite}
\usepackage{amsmath,amssymb,amsfonts}
\usepackage{algorithmic}
\usepackage{graphicx}
\usepackage{textcomp}
\usepackage{xcolor}
\usepackage[hidelinks]{hyperref}
\usepackage{enumitem}
\usepackage[accsupp]{axessibility}

\def\BibTeX{{\rm B\kern-.05em{\sc i\kern-.025em b}\kern-.08em
    T\kern-.1667em\lower.7ex\hbox{E}\kern-.125emX}}
\begin{document}

\title{Fidelity-preserving Learning-Based Image Compression: Loss Function and Subjective Evaluation Methodology\\
\thanks{This work is funded by FCT/MCTES through national funds and when applicable co-funded EU funds under the project DARING with reference PTDC/EEI-COM/7775/2020.}
}
\author{
   \IEEEauthorblockN{Shima Mohammadi\IEEEauthorrefmark{1}, Yaojun Wu\IEEEauthorrefmark{2}, Jo\~{a}o Ascenso\IEEEauthorrefmark{3}}
   \IEEEauthorblockA{\IEEEauthorrefmark{1}\IEEEauthorrefmark{3}Instituto Superior Técnico - Instituto de Telecomunicações}
   \IEEEauthorblockA{\IEEEauthorrefmark{2}Bytedance}
   \IEEEauthorblockA{\IEEEauthorrefmark{1}shima.mohammadi@lx.it.pt\IEEEauthorrefmark{2}wuyaojun@bytedance.com\IEEEauthorrefmark{3}joao.ascenso@lx.it.pt}
}

\maketitle

\begin{abstract}
Learning-based image compression methods have emerged as state-of-the-art, showcasing higher performance compared to conventional compression solutions. These data-driven approaches aim to learn the parameters of a neural network model through iterative training on large amounts of data. The optimization process typically involves minimizing the distortion between the decoded and the original ground truth images. This paper focuses on perceptual optimization of learning-based image compression solutions and proposes: i) novel loss function to be used during training and ii) novel subjective test methodology that aims to evaluate the decoded image fidelity. According to experimental results from the subjective test taken with the new methodology, the optimization procedure can enhance image quality for low-rates while offering no advantage for high-rates.
\end{abstract}


\section{Introduction}
Learning-based image codecs have emerged as state-of-the-art techniques for image compression. These codecs follow a similar pipeline to conventional codecs, which includes transform, quantization, and entropy coding. However, there are some key differences in the underlying coding engine. Instead of relying on traditional linear transformations, learning-based codecs are often implemented using convolutional neural networks (CNNs) and non-linear activation layers. Moreover, by learning the transformations directly from the data, these codecs can represent the underlying manifold of visual data, especially complex patterns and structures, very efficiently.

\par In learning-based image codecs, all the components are jointly trained in an iterative process which computes the model's parameters by minimizing some optimization target, often expressed in a rate-distortion loss function. While the naive approach is to optimize the codec by using Mean Square Error (MSE) as the distortion metric,
it is not always the best option, especially since this pixel-wise difference is not able to capture the perceived quality. To address this problem, more advanced perceptual loss functions have been proposed for learning-based image codecs which take into account perceptual factors such as structural similarity, texture, and contrast. Therefore, the model can be trained to prioritize the preservation of important visual details while discarding less relevant perceptual information. 

\par Generative Adversarial Networks (GANs) \cite{creswell2018generative} have achieved a significant success in several domains, including image restoration, fake media generation, and learning-based image compression. GAN-based image codecs often employ an architecture with a generator and a discriminator network and are able to reconstruct more visually pleasing images. However, GAN image coding models are famous to produce unique artifacts such as checkerboard patterns, jagged edges, color shifts, banding, texture replacement, and in the most severe case, images that can look forged due to visual inconsistencies. 

\par Moreover, an interplay between rate, distortion (also referred as \textit{fidelity}), and perception (also referred as \textit{appeal}) as defined in \cite{blau2019rethinking} must be considered when GAN-based methods are used for optimization of learning-based image codecs. For example, when the goal is to achieve high perceptual quality (\textit{appeal}), it usually requires higher rate or higher distortion, i.e. lack of \textit{fidelity}. Often, GAN-based image coding models prioritize image \textit{appeal} at the cost of some loss of \textit{fidelity}. The impact of that most objective quality metrics are not able to measure image quality reliably \cite{mentzer2020high}\cite{lao2022attentions} especially because their aim is to just measure \textit{fidelity} (or distortion as in \cite{blau2019rethinking}) and some subjective assessment methodologies, such as single or double stimulus, may also not be reliable, especially because their aim is to measure \textit{appeal} (or perception as in \cite{blau2019rethinking}). 

 
\par The main objective of this paper is to investigate the benefits of perceptual optimization of learning-based image coding. In this context, this paper has the following contributions:
\begin{enumerate}[leftmargin=*]
  \item Proposing a loss function and training procedure which aim to improve the perceptual quality of decoded images. This loss function is integrated into a learning-based image codec \cite{Bytedance_response} and includes a combination of the LPIPS (Learned Perceptual Image Patch Similarity) quality metric \cite{zhang2018perceptual} and an adversarial loss obtained with a GAN-trained neural network. This approach is particularly well suited to preserve \textit{fidelity} as much as possible, minimizing significant deviations in the decoded image texture or structure, which has often been observed in previous work.
  \item To evaluate the perceptual performance gains achieved by the perceptually optimized codecs, a novel subjective assessment methodology was proposed based on JPEG AIC Part 2 - Annex A \cite{JPEGAIC-2b}, that was originally developed for near-visually lossless quality assessment. This is a triplet subjective assessment test which aims to evaluate \textit{fidelity} loss and not only the \textit{appeal} of the images and thus, it is more suitable for image compression scenarios.
\end{enumerate}

The subjective assessment study involves two identical codecs with the same architecture but different models, one trained for classical quality metrics and the other trained to achieve improved perceptual quality by exploiting GAN adversarial loss. The subjective assessment test results including the decoded and reference images are available online \footnote{\href{https://github.com/shimamohammadi/vcip2023}{https://github.com/shimamohammadi/vcip2023}}. The experimental results allow to conclude for which type of content and for which test conditions (e.g. rate) GAN-based image compression can provide benefits.
\vspace{-2pt}
\section{Related Work}
\vspace{-2pt}
In many learning-based image compression solutions, metrics such as MSE or Multiscale structural similarity (MS-SSIM) \cite{Wang2003MSSIM} are used in the loss function and thus for optimization. However, in \cite{rippel2017real, agustsson2019generative, mentzer2020high}, the adversarial learning framework was leveraged to enhance the perceptual quality of the generated compressed images. The first work, by Rippel et al. \cite{rippel2017real}, proposed an autoencoder architecture featuring pyramidal decomposition, with bitplane division, adaptive arithmetic coding, and codelength regularization. Adversarial training was employed to achieve more visually pleasing reconstructions for very low bitrates. However, experimental results are only shown for objective quality metrics. Agustsson et al. \cite{agustsson2019generative} proposed a codec based on conditional GANs that operates on the full-resolution image and was trained in combination with a multi-scale discriminator. This approach achieve higher qualities at extremely low bitrates (below 0.1bpp) by synthesizing some parts of the image (in some cases guided by a semantic map). However, in this type of approach, some regions of the image may significantly deviate from the original image, lacking in terms of \textit{fidelity}. Mentzer et al. \cite{mentzer2020high} proposes a new codec architecture, studying the impact of normalization layers, generator and discriminator architectures, training strategies, and perceptual loss functions (including an adversarial loss). This approach delivers visually pleasing reconstructions that are perceptually similar to the input across a broad range of bitrates. A user study was made but using pairwise comparisons, not allowing to evaluate the \textit{fidelity} that may occur when the loss function is used. He et al. \cite{he2022po} proposes a learning-based codec with a new loss function, that includes terms that aim to enhance the perceptual quality: Charbonnier loss, LPIPS loss, hinge-form adversarial loss and style loss. They evaluate their codec using just objective quality metrics (and a few examples) and lacks the perceptual assessment of each loss in the final decoded image.

\par Mohammadi et al. \cite{shima2022PCS} has studied the perceptual impact of using several objective image quality metrics in the optimization process of learning-based codecs. Through a crowdsourcing pairwise subjective assessment test, it is shown that the choice of objective quality metric and the characteristics of the image content plays a crucial role in the final perceived image quality. However, this study does not include GAN adversarial losses. Sun et al. \cite{Sun2021} leverage the advantage of using both MSE and MS-SSIM in the loss function through an online loss function adaptation by reinforcement learning. In \cite{Sun2021}, the trade-off between PSNR and MS-SSIM is controlled to achieve better visual quality, as measured by the VMAF metric. Chen et al. \cite{Chen2021ProxIQA} proposed an alternative optimization strategy by introducing a proxy neural network as a surrogate for the non-differentiable perceptual quality metric (VMAF) \cite{rassool2017vmaf}. This proxy network acts as a perceptual model serving as a loss layer that is updated during training.

\par These works showcase the importance of leveraging the strengths of multiple quality metrics in the loss function and the use of the adversarial loss to enhance perceptual quality. However, the performance gains achieved by such perceptual optimization are not reliably subjectively assessed, especially in terms of \textit{fidelity}.

\vspace{-2pt}
\section{Perceptually Optimized Learning-based\\Image Codec}
\vspace{-2pt}
The perceptual optimized loss function will be integrated into the learning-based image codec of \cite{Bytedance_response}. This image codec was proposed within context of the call for proposals of the JPEG AI ad-hoc group. This solution proposes a decoupled framework where the entropy decoding process is independent of the latent reconstruction process enabling massive parallelization, which leads to significant decoding time savings while still achieving higher decoding qualities. Moreover, wavefront processing is introduced in the auto-regressive model, where multiple rows can be simultaneously processed. In addition, device interoperability is achieved by the design of a neural network model quantization process. A series of coding tools to further improve the coding efficiency, reduce complexity and perform rate adaptation are introduced, namely adaptive quantization, latent refiner, tiling and down-sampling/up-sampling filters.

\par
\begin{figure}[h]
    \captionsetup{skip=3pt}
    \centering
    \includegraphics[scale=0.3]{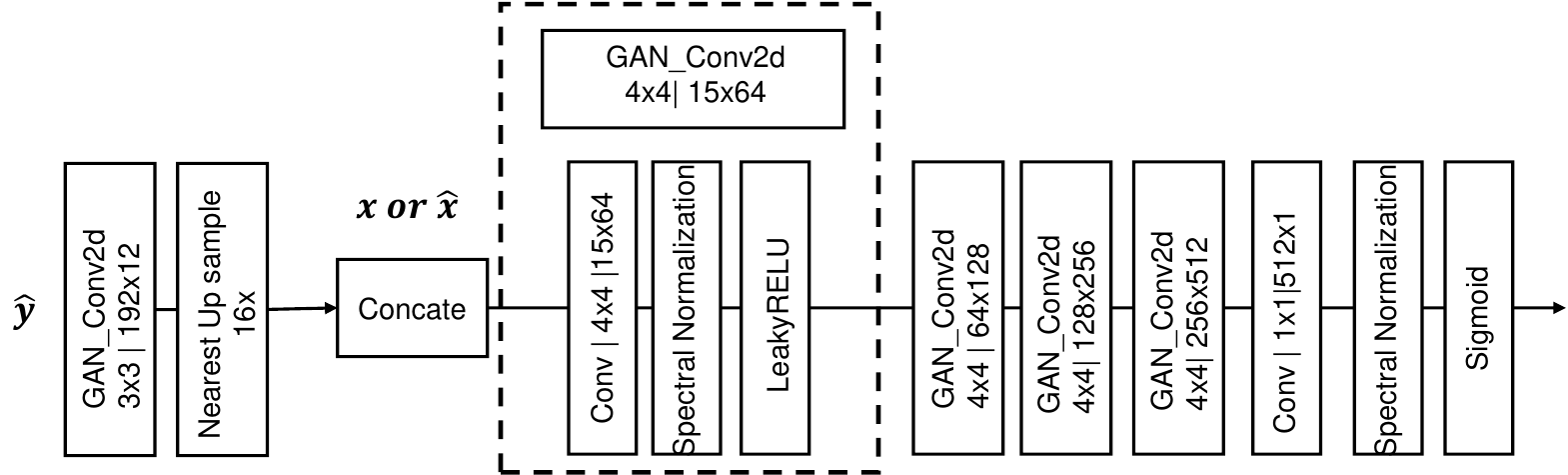}
    \caption{Discriminator architecture.}
    \vspace{-10pt}
    \label{fig:discriminator}
\end{figure}

 During the training procedure, two image compression models (and thus codecs) can be obtained, which only differ on the parameters used:
\begin{itemize}[leftmargin=*]
    \item LBIC-CO: learning-based image coding conventional optimized solution where 5 models are initially trained using the loss function of (\ref{equ:NoGAN}) which includes MSE and luma MS-SSIM image quality metrics. The 5 models are also fine-tuned to train 11 models for rate matching, allowing a finer control. 
    \begin{align}
        \small \mathcal{L} = \lambda(QP)[\alpha\textrm{MSE} + \beta(1-\textrm{MS-SSIM}_{Y})] + R
    \label{equ:NoGAN}
    \end{align}
    with $\mathcal{L}$ as the loss, $R$ as the rate of the latent and hyper-prior and MSE/MS-SSIM as the quality metrics. Based on compressAI models \cite{compressai}, it was fitted a linear model to the $\lambda_{\textrm{MS-SSIM}}$ and $\lambda_{\textrm{MSE}}$ and the following relationship was found: $\lambda_{\textrm{MS-SSIM}} = 1275\lambda_{\textrm{MSE}}$. Therefore, $\alpha=255^2$ as in compressAI and $\beta=1275$. This represents the best combination of MSE/MS-SSIM based on the much popular compressAI framework.
    \item LBIC-PO: proposed learning-based image coding perceptually optimized solution which was designed to maintain a high level of \textit{fidelity}. In this case, additional fine-tuning is  carried out over LBIC-CO models, where the loss function is replaced with (\ref{equ:GAN}), that now also includes adversarial loss and the LPIPS quality metric. 
    \begin{flalign}\label{equ:GAN} 
         \small \mathcal{L} = & \zeta\lambda(QP)[\alpha(\eta\textrm{MSE} + \theta G_{a}) + \rho\textrm{LPIPS} &\\\nonumber
         & + \sigma(1-\textrm{MS-SSIM}_Y)] + R &
    \end{flalign}
    where $G_{a}$ is the adversarial loss of the discriminator. Note that MSE and MS-SSIM quality metrics are used to maintain a high-level of \textit{fidelity} between the decoded and original images. To provide the best trade-off between rate, \textit{fidelity} and \textit{appeal}, the hyperparameters $\zeta=\frac{5}{6}$, $\eta=\frac{3}{8}$, $\theta=0.75\times10^{-4}$, $\rho=0.005$, $\sigma=0.5$ were manually adjusted to weight each quality metric and the adversarial loss in a suitable way. 
\end{itemize}
In the discriminator, the latent representation $\hat{y}$ (after entropy decoding) serves as the conditional input, the original YUV $x$ and reconstructed YUV $\hat{x}$ are used by the discriminator to test whether the input is real (original image) or fake (distorted image), respectively. This discriminator is only used during training and allows to find a perceptual-based model that reconstructs images as closely as possible to the original image. The discriminator architecture is shown in Fig. \ref{fig:discriminator}. 
\vspace{-1.5em}
\vspace{-2pt}
\section{Subjective Evaluation Methodology}
\vspace{-2pt}
In this section, a novel subjective test methodology is proposed based on JPEG AIC-2 Part A and can be classified as triplet comparison without forced choice \cite{men2021subjective}. This methodology allows to achieve high accuracy and low bias since the subjects decision is very straightforward, doesn't require training and is robust to changes in viewing conditions, especially in comparison with single and double stimulus methodologies (which require to score the image quality in a predetermined scale). Moreover, small quality differences can be detected by the subjects, which is very important for this assessment since the decoded images may appear rather similar (in some cases).

\par Since the aim is to evaluate \textit{fidelity}, three stimuli (reference and two distorted images), are presented: the original image is positioned at the center, and the two decoded images obtained with the codecs described in the previous section is placed on the right or left. The decoded images have the same bitrate to enable a fair comparison. This is much preferable to a pairwise comparison, where the original image is not present or a double stimulus methodology where the two decoded images are not shown simultaneously (and thus focus on \textit{appeal}) as seen in past performance assessment studies.

\par During both the training and the testing phases, participants were forced to use full screen mode. Subjects were a mix of experts and non-experts and used an internet browser to conduct the subjective test. Each subject was instructed to carefully examine the three provided images (original at the center and the two decoded) and then decide which one of the two images on the left or right is more similar to the reference image in the middle. The images are shown in their original resolution without any scaling. If subjects couldn't detect differences between the two coded images or had no preference between them, they are allowed to select ``No preference between A and B'' button. To ensure no bias, the triplets were randomly shuffled for each subject. Besides, the location (left or right) of the decoded images of the two codecs under evaluation was also randomly chosen. 
\vspace{-2pt}
\section{Performance Evaluation}
\vspace{-2pt}
This section describes the evaluation of the proposed perceptually optimized learning-based image codec with the novel subjective assessment methodology.

\subsection{Subjective Test Platform}
\par
\begin{figure}
    \captionsetup{justification=centering}
    \centering
    \includegraphics[scale=0.12]{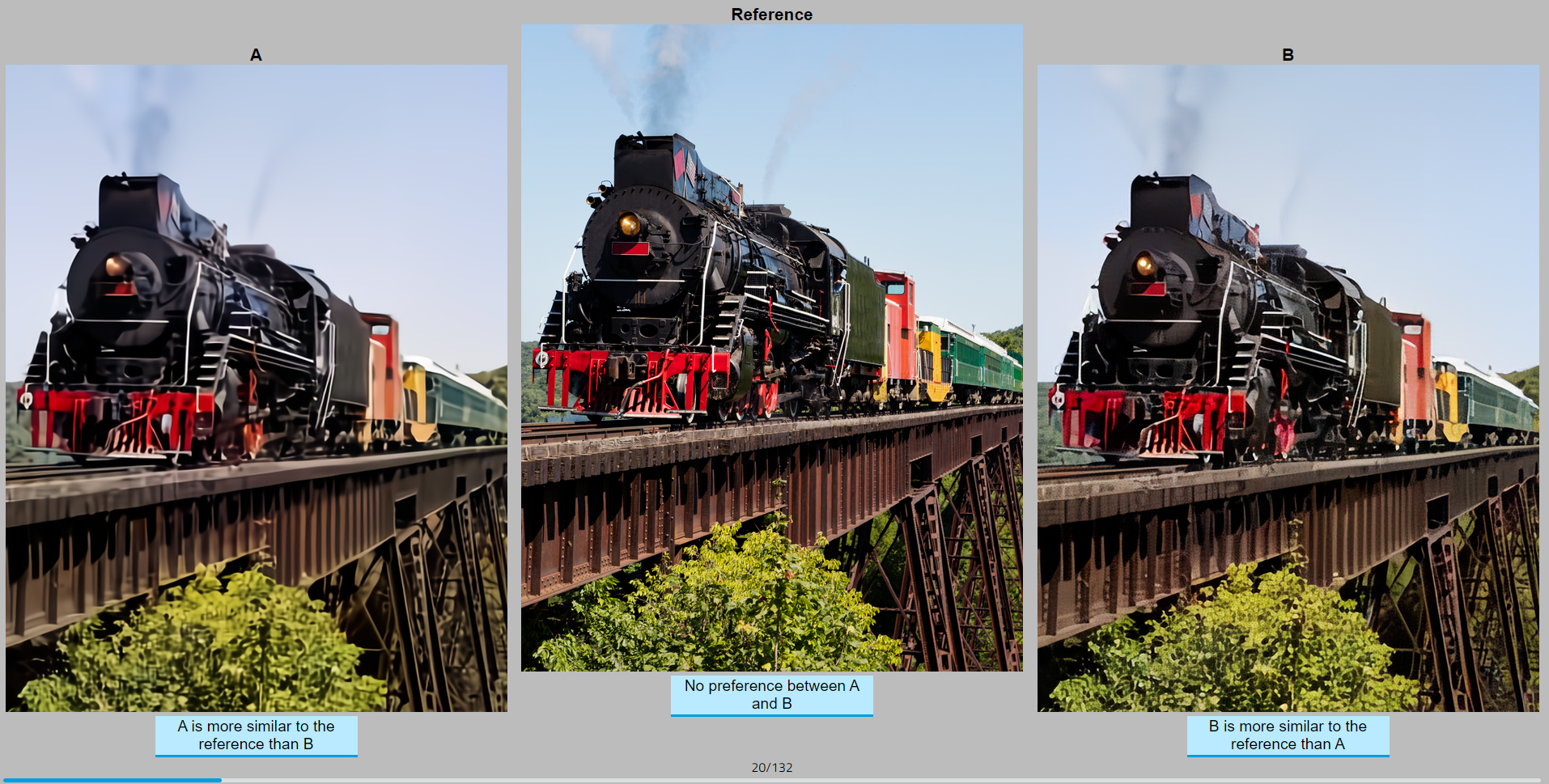}
    \caption{Crowdsourcing platform layout for the triplet comparison subjective assessment test.}
    \label{fig:platform}
    \vspace{-5mm}
\end{figure}

 To conduct the proposed triplet comparison subjective test, a web-based crowdsourcing platform was employed with NodeJS platform and MongoDB database \cite{platform}. The platform was designed to detect the monitor resolution, and only allows the subjects who have minimum resolution of fullHD ($1920\times1080$) and a display size of 13 inches to participate in the test. Prior to starting the subjective test, subjects were required to fill out a form with data such as their name, email, age, gender, and display size. The subjective test begins with a training phase with easy to answer trials to familiarize the subjects with the test objective as well as the platform itself. Upon completing the training phase, subjects begin the test phase. Fig. \ref{fig:platform} shows the layout of the crowdsourcing web platform. 

\subsection{Test Material}
To evaluate the codecs under test, 46 reference images of the JPEG AI dataset were coded with both LBIC-CO and LBIC-PO codecs at the five bitrates 0.06, 0.12, 0.25, 0.5, 0.75 bpp as defined in the JPEG AI common training and test conditions \cite{JPEG_AI_train_test_condition}. The coded images were cropped to fit the side-by-side layout of the crowdsourcing platform considering a minimum display resolution of ($1920\times1080$) which was enforced during the subjective test. The cropped size is selected to be $620\times800$ to fit the three images, the two coded images under test plus the reference image, into the screen with the selected resolution. Moreover, the location of the crop region was carefully selected to cover a salient region of the image. 


\subsection{Triplet Selection}
To conduct a triplet comparison subjective test with all decoded images for all bitrates, 320 ($46\times5$) triplets are required where 46 is the number of reference images and 5 is the number of target bitrates. However, a subjective test with 320 triplets is impractical as it would be very lengthy and could lead to subject fatigue. Therefore, reducing the number of triplets becomes essential. The idea is to sub-sample the triplets to obtain a meaningful subset which is representative of the whole set. One way to reduce the number of triplets is to filter out cases where subjects select ``No preference''. This typically happens when two decoded images are very close in quality, and the user doesn't have a strong preference. Minimizing these cases helps to obtain a smaller set of triplets.The used procedure is defined as follows:
\begin{enumerate}[leftmargin=*]
    \item First a preliminary subjective test was conducted with a few experts. The platform was the same as previously described. Five subjects, 3 males, and 2 females participated in the test who labeled all the triplets. Each triplet consists of LBIC-CO and LBIC-PO coded images using the same reference as well as the same bitrate.
    \item The PSNR image quality metric was used to evaluate the similarity between the decoded images obtained by LBIC-CO and LBIC-PO within a triplet.
    \item A triplet is removed if the PSNR computed in the previous step exceeds a predefined threshold $t$.
    \item Then, the no-preference classification rate denoted as $CR(t)={|S \cap P(t)|}\mathbin{/}{S}$ is computed, which accounts for the cases where subjects agree with the applied threshold. In this case, $S$ is the set of triplets where the ``no preference'' label was assigned in the preliminary test and $P(t)$ is the set of triplets where the PSNR is above the threshold (i.e. decoded images are visually similar). $||$ represents the cardinality of the set.
\end{enumerate}

\begin{figure}[ht]
\captionsetup{skip=0pt}
\centerline{
    \begin{tabular}{@{}c@{}}
         {\includegraphics[scale=0.38]{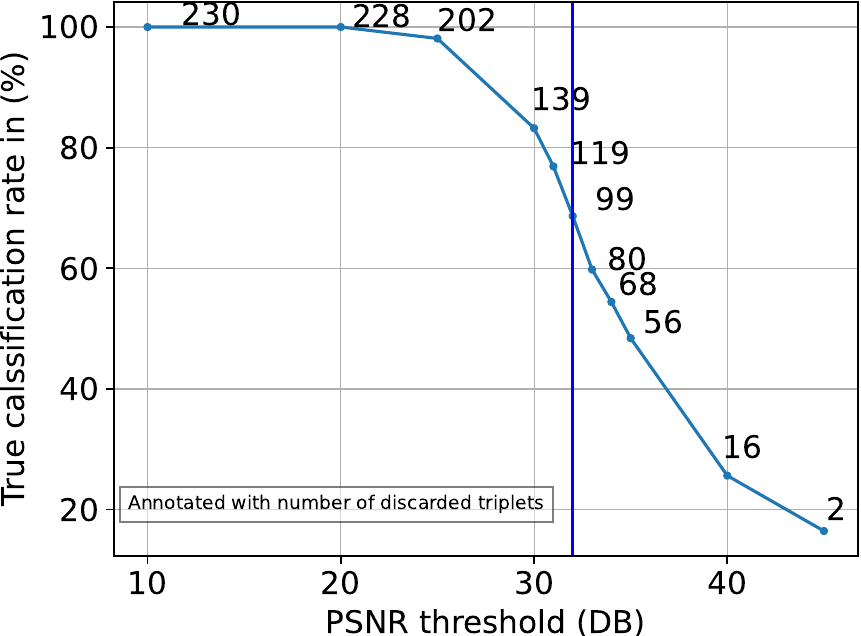}} 
    \end{tabular}
   }
\caption{True classification rate $TCR$ }
\label{fig:Th_setting}
\vspace{3pt}
\centerline{   
   \begin{tabular}{@{}c@{}}
        {\includegraphics[scale=0.35]{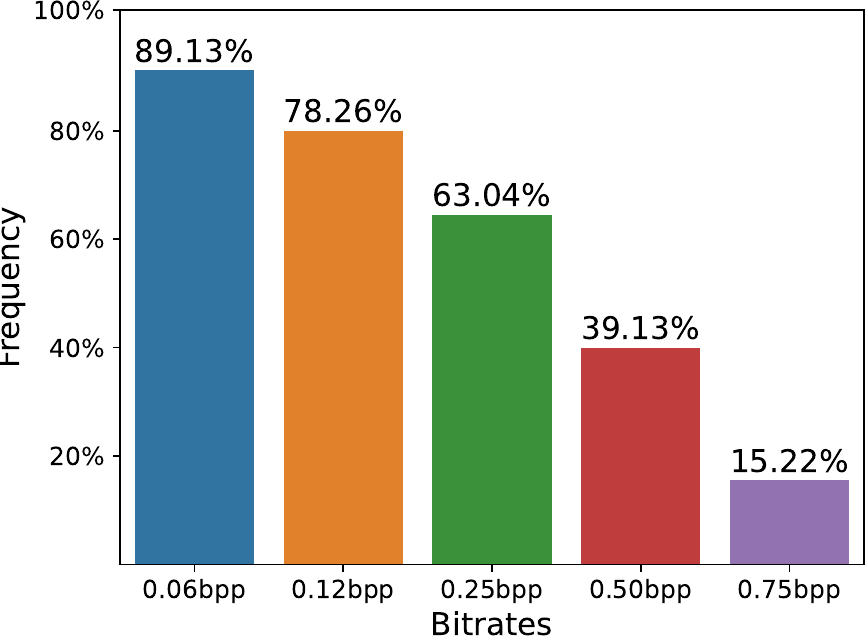}} 
   \end{tabular}}
\caption{Selected triplets (\%) for each target rate}
\vspace{-1.5em}
\label{fig:available_bitrates}
\end{figure}

The results of the above procedure for different threshold values is shown in Fig. \ref{fig:Th_setting} where it is annotated with number of removed triplets for a range of thresholds. As this figure suggests, by setting threshold of 10dB, all the triplets are removed while a threshold of 45dB is equivalent of removing only 2 triplets. In this case, a threshold of 32dB was chosen, as it removes 99 triplets and keeps 131 triplets to be included in the subjective test. Thus results in a subjective test with a reasonable duration. The $CR$ for this threshold is $\approx 70\%$, which allows for valid conclusions to be taken.
\vspace{-1.0em}

The number of triplets that were selected to be evaluated in the subjective experiment (not discarded with the aforementioned procedure) is depicted in Fig. \ref{fig:available_bitrates} according to the bitrate. As shown, after applying the threshold approximately 90, 78, 64, 40, and 15 \% of the triplets remains in 0.06, 0.12, 0.25, 0.50, 0.75 bpp respectively. This was expected since for high bitrates, the decoded images within each triplet are very close to the original image and between themselves and thus do not need to be selected.
\vspace{-5pt}
\subsection{Experimental Results}
This section is dedicated to the experimental results of the final subjective test. 

\subsubsection{Subjective Data Analysis}
Overall, 20 subjects, 13 male and 7 females, were invited to the crowdsourcing subjective test using the platform previously described. The subject’s age is distributed between 23 and 53 with an average of 34. The monitor resolution of 12 of the subjects was $1920\times1080$, and the remaining 8 was ($2560\times1440$). 

\subsubsection{Analysis per Bitrate}
\begin{figure}
\captionsetup{skip=0pt}
    \centering
    \includegraphics[scale=0.39]{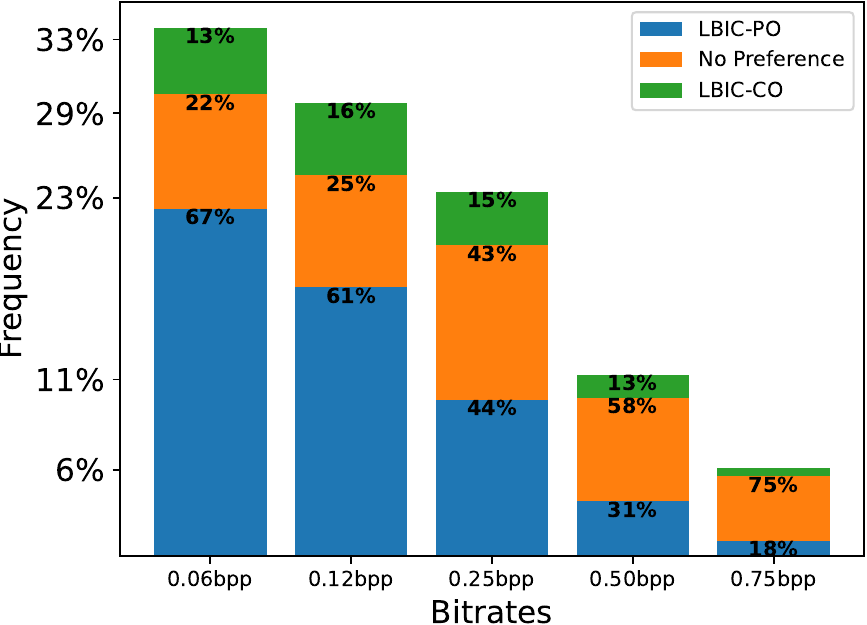}
    \caption{Distribution of votes per bitrate.}
    \label{fig:bitrate}
    \vspace{-5mm}
\end{figure}

The experimental results of the subjective test for the available five bitrates is shown in Fig. \ref{fig:bitrate}, where each bar is representative of the total number of triplets evaluated for each bitrate, considering the total number of triplets as in Fig. \ref{fig:Th_setting}. Moreover, each bar is divided into three parts of different color, which represent the percentage of votes for LBIC-CO, LBIC-PO and ``No preference'' based on the triplets that were subjectively assessed.

\par Overall, LBIC-PO coded images were in most cases considered as closer to the reference than LBIC-CO coded images in all the five bitrates. In the 0.06 and 0.12 bpp, LBIC-PO coded images have received votes more than 50\% of the times. However, for 0.25 bpp, the votes for LBIC-PO coded images and ``no preference'' between the two codecs is almost the same. For 0.5 and 0.75 bpp, the majority of subjects had no preference between LBIC-CO and LBIC-PO images with 58 and 75 percent respectively. This allows to conclude that for these higher bitrates there is not much difference between the two decoded images under evaluation and thus a model trained with adversarial loss and the LPIPS quality metric is not advantageous and may even be detrimental. 

\subsubsection{Analysis per Content}
\begin{figure}
\captionsetup{skip=0pt}
    \centering
    \includegraphics[scale=0.24]{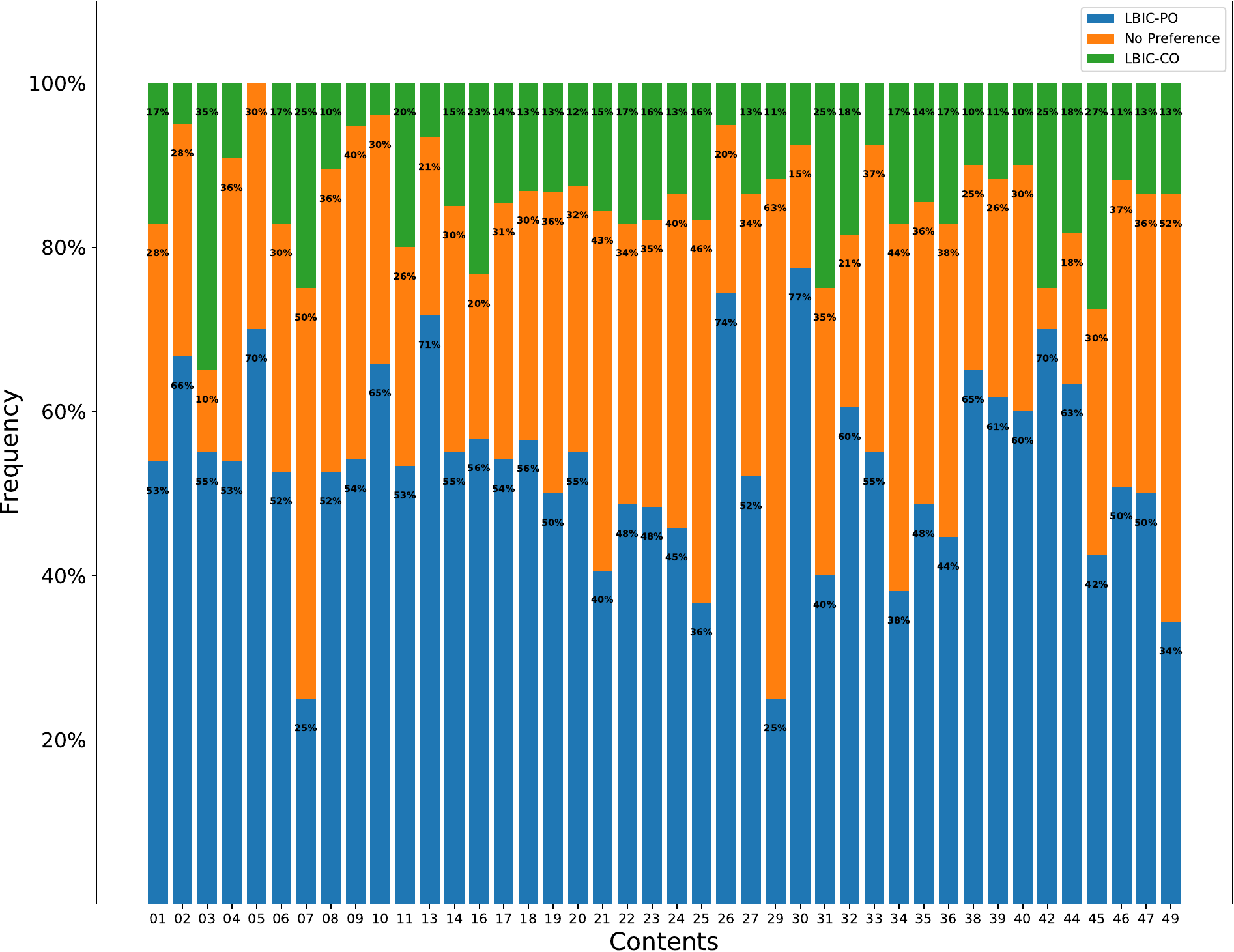}
    \caption{Distribution of votes per reference image.}
    \label{fig:content}
   \vspace{-18pt}
\end{figure}

The subjective test results are also analyzed for every selected test image and are shown in Fig. \ref{fig:content}. This allows to understand any preference or discrepancy considering every test image under evaluation. In this figure, each bar is representative of a reference and is annotated with the number of votes for every possible choice by the subjects: LBIC-CO, LBIC-PO and ``No preference''. In this case, the bar height goes to 100\% for every test image, and thus is only considered the triplets subjectively evaluated; this allows to clearly visualize the differences in the subject's opinion. 
For most references, the LBIC-PO coded images were preferred over the LBIC-CO coded images for more than 50\% of the time except for image \#7, \#21, \#22, \#23, \#24, \#25, \#29, \#34, \#46, \#45 and \#49 where it was less than 50\% but higher than 25\%. In the case of image \#7, half of the time, there was no preference of one over the other, while in the other half the preference between the two codecs were equal and it was at 25\%. However, there is not a single case where LBIC-CO is better than LBIC-PO. This allows to conclude that the proposed loss function used for optimization is reliable and can be used for a wide range of images, although for some cases may not provide significant benefits. A qualitative analysis of some selected images will be made available at the Github repository.
\vspace{-2pt}
\section{Conclusions}
\vspace{-2pt}
This paper proposes a new \textit{fidelity}-preserving perceptual optimization procedure for learning-based image compression with an adversarial loss, the LPIPS quality metric and a new subjective assessment methodology to evaluate the \textit{fidelity} gains. The subjective test results clearly show that the advantages of the perceptually optimized image codec since subjects consistently selected for a wide range of images, the corresponding decoded images as more similar to the reference images, especially for the lowest rates. As future work, the plan is to include this type of approach in the JPEG AI verification model at the encoder side (which is non-normative) while keeping the decoder fixed, thus allowing to select the model dynamically at the encoder depending of the test content and the target bitrate.


\bibliographystyle{ieeetr}
\bibliography{IEEEexample}

\end{document}